\documentclass[a4paper, 12pt]{report}
\usepackage{graphicx}
\usepackage{amsmath, amsthm, amsfonts}
\begin{document}
\title{Entropy in the Thermal Model}
\author{Maciej Andrzej Stankiewicz}
\date{October 2003}
\maketitle

\begin{abstract}
A brief review of the Hadron Gas model with reference to high
energy heavy ion collisions is presented.   The entropy dependence
on temperature and baryonic chemical potential is numerically
calculated, together with the entropy distribution between baryons
and mesons. The theoretical entropy for a QGP with equivalent
parameters is also calculated.  It is shown that at low
temperatures the dominant entropy contribution comes from baryons,
while at high energies the dominant contribution comes from
mesons.  The turnover from baryon to meson domination occurs at $T
\approx 140 \hbox{MeV}$, which corresponds to an energy domain of
the AGS at Brookhaven National Laboratory.
\end{abstract}

\tableofcontents


\chapter{Introduction}

Relativistic heavy ion collisions provide us with the opportunity
to study strongly interacting matter at extremely high temperature
and pressure.  It is the current goal of collision experiments to
attempt to create a new state of matter known as a Quark-Gluon
Plasma (QGP).  This refers to a state where the quarks and gluons,
which were originally bound into nucleons, become deconfined and
form a self-consistent plasma.  This form of matter is believed to
have existed in the very early universe (shortly after the Big
Bang).

The Quark-Gluon Plasma was originally proposed as a byproduct of
QCD. Quantum chromodynamics was developed as a theory of the
strong interaction, but unlike QED, it is almost impossible to
obtain analytic solutions.  Lattice QCD (discrete points represent
space-time) has been worked on numerically, and it suggests that a
QGP forms when the temperature ($\sim 150-200 \hbox{ MeV}$) or
density (up to 5-10$\rho_0$, $\rho_0 \sim 3\times
10^{14}\hbox{g/cm}^3$, the nuclear density) becomes large enough.

The search for the QGP is currently ongoing.  As the beam energies
of heavy ion accelerators continue to increase, the systems
resultant from the collisions have initial temperatures and
densities close to the above thresholds.

Since direct observation of the QGP is impossible (it is very
short-lived, and its component quarks and gluons become confined
into hadrons before they reach the detectors), one has to infer
its formation from the observed distributions of hadrons, leptons
and photons.  This is highly non-trivial, as most hadronic
observables reflect conditions at freeze-out (when the hadrons
stopped interacting -- similar to the surface of last
scattering.), by which time final state interactions may have
caused loss of all information about the initial partonic state.

After the release of the latest data from RHIC, there has been
some speculation at to whether data showed signs of existence of
QGP, but the issue has not been settled yet.  It is hoped that
with still higher energies expected in the near future at LHC,
results signalling the formation of QGP will be observed.

\section{Ideal Hadron Gas Model}

While QGP is expected to be the high-energy limit of a collection
of hadronic particles, at lower temperatures there is a different
model.  The conventional, confined phase is usually referred to as
Hadronic Gas (HG), consisting of hadrons of different types
(including the short-lived resonances) such as $\pi$, $\rho$, $N$,
$\Delta$, $K$, whose properties (mass, spin, degeneracy) are well
known.

This phase does exist in heavy ion collisions. Even if a QGP is
created, it will rapidly expand until the temperature drops below
the critical value, and hadronization takes place.  This means
that the deconfined quarks and gluons of the QGP will form bound
states of baryons, mesons and anti-baryons.

The system will then form a small volume that is filled with a
huge variety of interacting hadrons.  Assuming that a state of
equilibrium is reached it seems plausible to treat the system
(fireball) using the methods of statistical mechanics.  The
Thermal Model can be applied to a system at thermal equilibrium,
although to find the particle multiplicities, chemical equilibrium
also required.

\subsection{Thermal and chemical freeze-out}

Directly after the hadronization, the particles will be
sufficiently close to interact strongly (distances of order
$1\hbox{fm}$), and it is expected that this form of interaction
will cause the system to reach a state of ``chemical equilibrium''
(chemical referring to the composition of the fireball). Hence,
chemical equilibrium means that the number of each form of
particle will be constant - the rate of creation/annihilation for
each type of particle will be exactly equal.  This will then
continue until the gas cools further (it is expanding), and the
temperature drops below a critical value. This is called
\emph{chemical freeze-out} - the point at which inelastic
collisions cease and all particle ratios are frozen.

A second form of freeze-out, known as \emph{thermal freeze-out} is
often also introduced.  After the chemical freeze-out, once
particle ratios become fixed, particles still interact elastically
(eg. $\pi + N \rightarrow \Delta \rightarrow \pi  + N$).  These
types of reactions have larger cross-sections and continue to
occur, redistributing momentum in the system while unchanging the
chemical composition.  The fireball continues expanding and
cooling until these interactions cease (thermal freeze-out). The
particles then fly off towards the detectors.

By observing the particle abundances (which show the properties of
the system at chemical freeze-out), one finds that (at CERN SPS
energies) the distribution is characteristic of one at $T \sim 170
\hbox{ MeV}$. By observing the particle momentum distributions
(which characterize the system at thermal freeze-out), one can
infer that (for CERN SPS), the  thermal freeze-out occurs later at
$T \sim 130 \hbox{ MeV}$.

\newpage

\section{Theoretical Formulation of the Thermal Model}

In the fireball, the number of particles is not fixed, and so one
works with a Grand Canonical Ensemble.  Firstly, consider the
analysis of a \emph{static} fireball.  Hence the system lives for
a period of time in a fixed volume $V$.  Then for one type of
particle:
\begin{equation}
\Omega_{GC} = -kT \ln Z
\end{equation}
where $Z$ is the partition function, the form of which depends on
whether the particle follows Bose-Einstein or Fermi-Dirac
statistics. The specific quantities will be discussed in detail in
the next section.  The multiplicity of hadrons of species $i$ is
given as:
\begin{equation}
N_i = Vn_i = \frac{Vg_i}{(2\pi\hbar)^3} \int f_i(p) d^3p
\end{equation}
where $n_i$ is the number density, $g_i = 2J_i + 1$ is the spin
degeneracy factor, and $f_i(p)$ is the momentum distribution
function. In a thermodynamic equilibrium distribution, these
functions have a relatively simple form:
\begin{equation}
f_i(p)
= \left[ e^{\frac{E_i(p)-\mu_i}{kT}} + \epsilon \right]^{-1}
= \left[e^{\frac{\sqrt{p^2+m_i^2}-\mu_i}{kT}} + \epsilon \right]^{-1}
\end{equation}
Here $k$ is Boltzmann's constant, $T$ is the temperature of the
system, and $\mu_i$ is the chemical potential of hadron $i$.  The
quantity $\epsilon$ equals $+1$ for fermion (FD statistics), and
$-1$ for bosons (BE statistics). The limit $\epsilon \rightarrow
0$ corresponds to classical (Boltzmann) statistics.

The chemical potential $\mu_i$ is what ultimately distinguishes
the hadrons from each other.  The quantities which are conserved
for particle interactions within a hadron gas (which is
sufficiently short-lived and short-range that it only interacts
strongly) are the baryon number, charge and strangeness.  The
chemical potential is then a linear combination of the three
potentials:
\begin{equation}
\label{Chemical}
\mu_i = \mu_B \verb"B"_i + \mu_Q \verb"Q"_i + \mu_S \verb"S"_i
\end{equation}
Here, $\verb"B"_i$, $\verb"Q"_i$ and $\verb"S"_i$ are the baryon
number, charge and strangeness of the $i$'th hadron respectively.
This can be extended by adding the charm and bottomness potentials
in a similar linear fashion.  However, for the current
calculations, only the particles composed of the $u$, $d$ and $s$
quarks have been used.

Introduction of the chemical potentials $\mu_B$, $\mu_Q$ and
$\mu_S$ allows us to fulfill the appropriate conservation laws
(for strong interactions).  The strangeness of the system must be
zero:
\begin{equation}
\label{NonStrange} \sum_i \texttt{S}_i N_i = V \sum_i \texttt{S}_i n_i = 0
\end{equation}
This can be implemented without knowing anything about the volume
$V$.

\newpage

Next, the total charge of the fireball must be the same as the
total charge of the colliding nuclei.  This calculation is however
complex (volume modulated), but one needs to observe that the same
factor exists in total baryon number.  Dividing, we require the
electric charge to baryon ratio to be:
\begin{equation}
  \left(\frac{\sum_i \texttt{Q}_iN_i}{\sum_i \texttt{B}_iN_i} \right)_{HG}
= \left(\frac{\sum_i \texttt{Q}_iN_i}{\sum_i \texttt{B}_iN_i} \right)_{nuclei} =
\frac ZA
\end{equation}
This can be implemented for a general collision of two nuclei by
varying $\mu_Q$ until the above equation is satisfied.  For Pb+Pb
and Au+Au collisions, the ratios differ very slightly, but can be
very well approximated by:
\begin{equation}
\frac {N_{p}}{N_{n}} \approx 0.66 \qquad \hbox{ and } \qquad
\frac ZA \approx 0.40
\end{equation}
reflecting that there are slightly more neutrons than protons.  A
simplification I was encouraged to make was to to fix the
potential $\mu_Q = 0$.  As $m_p \simeq m_n$ to very good accuracy,
setting the electric charge potential to zero introduces a
charge-independence into the system, and as protons and neutrons
have almost the same mass, the model will produce an equal number
of them.  Hence $\frac ZA = 0.5$ for $\mu_Q = 0$.

This may seem like an unnecessary simplification, and a deviation
from real heavy ion collisions.  However this is still exact for
O+O and S+S collisions and it does produce a rather large
simplification in the working.  As it is, provided with $T$ and
$\mu_B$, there is only one unknown parameter $(\mu_S)$, and
although there is no analytic method to solve for $\mu_S$ to
satisfy (1.5), one can do it iteratively.  Without specifying
$\mu_Q$, there are two unknown parameters, and to satisfy both
strangeness and charge conservation requires iterative methods in
2-D.

If one does not implement this simplification and for a given $T$
and $\mu_B$ does actually work out the value of $\mu_Q$ for a
heavy ion collision, one finds that it is usually very small and
negative ($\mu_Q < 0$, to make neutrons more favourable than
protons). Typical values are  $-10\hbox{ MeV} < \mu_Q <-4 \hbox{
MeV}$, depending on the temperature.

Furthermore, the actual number of particles $N_i$ and entropy
$S_i$ depend on the volume $V$ of the fireball, and for this
project that is an unknown quantity.  I therefore always worked
with the number and entropy densities $n_i$ and $s_i$. However all
the particle proportionalities are still known.

The above is true for a static fireball - one that occupies a
fixed volume $V$.  In a more general case, when the expansion of
the system at freeze-out cannot be neglected, one needs to use the
more complex Cooper-Frye formula to calculate the total yield of
particles.  However it has been shown \cite{Mariusz} that in this
more complex system, the particle ratios stay the same as in the
static fireball case, as long as the thermodynamic parameters are
constant along the freeze-out surface, so to get the particle
ratios, one can work with a static fireball.


\chapter{Numerical Integration}

The hadrons come in two varieties: baryons (anti-baryons) which
obey Fermi-Dirac statistics, and mesons which obey Bose-Einstein
statistics.  Although the two behaviours are very different at low
temperatures, the formalism is very similar -- often just a change
of sign.  I'll thus deal with both cases using $\pm$ and $\mp$
notation, with $^{FD}_{BE}$ convention (top sign for fermions).

\section{Quantum Statistics Integrals}

\subsection{Number densities}
As mentioned in the previous section, we define a momentum
distribution function:
\begin{equation}
f_i = f_i(p) = \left[e^{\frac{E-\mu_i}{kT}} \pm 1 \right]^{-1} =
\left[e^{\frac{\sqrt{p^2+m_i^2}-\mu_i}{kT}} \pm 1 \right]^{-1}
\end{equation}
Then the number density of hadron $i$ is given by:
\begin{equation}
n_i = g_i \int \frac {d^3p}{(2\pi\hbar)^3}f_i(p)
\end{equation}
Assuming that the fireball is isotropic, the integral separates into angular and spatial parts:
\begin{equation}
n_i = g_i\int_0^{2\pi} d\phi \int_{-1}^{1} d(\cos \theta)
\int_0^\infty \frac {p^2 dp}{(2\pi\hbar)^3}f_i(p) =
\frac{g_i}{2\pi^2\hbar^3} \int_0^\infty f_i(p) p^2 dp
\end{equation}
Going over to a system of units where $\hbar = 1$, the
multiplicity of hadron $i$ becomes:
\begin{equation}
n_i = \frac {g_i} {2\pi^2} \int_0^\infty f_i(p) p^2 dp
= \frac {g_i} {2\pi^2} \int_0^\infty \frac {p^2 dp} {e^{\frac{E-\mu_i}{kT}}\pm 1}
\end{equation}
\newpage
\subsection{Entropy densities}

For a set Grand Canonical ensemble of particles $i$ in volume $V$, the partition function obeys:
\begin{equation}
\ln Z = \pm V g_i \int \frac{d^3p}{(2\pi\hbar)^3} \ln\left(1 \pm e^{-\frac{E-\mu_i}{kT}}\right)
\end{equation}
Then the entropy is defined by:
\begin{eqnarray*}
S_i
&= &\frac{\partial}{\partial T}(kT \ln Z) \\
&= &k \ln Z + \frac{kT}{Z}  \frac{\partial Z}{\partial T}  \\
&= &k \ln Z + \frac{kT}{Z}\cdot Z \frac{\partial}{\partial T} \left( \pm V g_i \int \frac{d^3p}{(2\pi\hbar)^3} \ln\left(1 \pm e^{-\frac{E-\mu_i}{kT}} \right) \right) \\
&= &k \ln Z \pm kTVg_i \int \frac{d^3p}{(2\pi\hbar)^3} \frac{\partial}{\partial T}\left[ \ln\left(1 \pm e^{-\frac{E-\mu_i}{kT}} \right)\right] \\
&= &\pm kVg_i \int \frac{d^3p}{(2\pi\hbar)^3} \left[\ln\left(1 \pm e^{-\frac{E-\mu_i}{kT}}\right)  + T\frac{\pm e^{-\frac{E-\mu_i}{kT}}\left( +\frac{E-\mu_i}{kT^2}\right)} {1 \pm e^{-\frac{E-\mu_i}{kT}}} \right] \\
&= &\pm kVg_i \int \frac{d^3p}{(2\pi\hbar)^3} \left[\ln \left(\frac{e^{\frac{E-\mu_i}{kT}}\pm 1}{e^{\frac{E-\mu_i}{kT}}}\right)  \pm  \left(\frac{E-\mu_i}{kT}\right) \frac{1} {e^{\frac{E-\mu_i}{kT}}\pm1 }  \right] \\
&= &\pm kVg_i \int \frac{d^3p}{(2\pi\hbar)^3} \left[ \ln\left(e^{-\frac{E-\mu_i}{kT}}\right) + \ln \left({e^{\frac{E-\mu_i}{kT}}\pm1}\right) \pm \left(\frac{E-\mu_i}{kT}\right) \frac{1} {e^{\frac{E-\mu_i}{kT}}\pm1 }  \right] \\
&= &\pm kVg_i \int \frac{d^3p}{(2\pi\hbar)^3} \left[ -\left(\frac{E-\mu_i}{kT}\right) - \ln f_i \pm \left(\frac{E-\mu_i}{kT}\right) f_i  \right] \\
&= &\pm kVg_i \int \frac{d^3p}{(2\pi\hbar)^3} \left[ -\ln f_i +  \left(\frac{E-\mu_i}{kT}\right) (-1 \pm f_i)  \right] \\
&= &\pm kVg_i \int \frac{d^3p}{(2\pi\hbar)^3} \left[ -\ln f_i +  \ln\left(\frac{e^{\frac{E-\mu_i}{kT}}}{e^{\frac{E-\mu_i}{kT}}\pm1}\left({e^{\frac{E-\mu_i}{kT}}\pm1}\right)\right) (-1 \pm f_i)  \right] \\
&= &\pm kVg_i \int \frac{d^3p}{(2\pi\hbar)^3} \left[ -\ln f_i +  \ln\left( \frac{(1 \mp f_i)}{ f_i} \right) (-1 \pm f_i)  \right] \\
&= &\pm kVg_i \int \frac{d^3p}{(2\pi\hbar)^3} \left[ -\ln f_i - \ln( 1 \mp f_i) (1 \mp f_i) + \ln(f_i)( 1 \mp f_i)\right] \\
&= &\pm kVg_i \int \frac{d^3p}{(2\pi\hbar)^3} \left[ \mp f_i \ln f_i - \ln( 1 \mp f_i) (1 \mp f_i) \right] \\
&= &    kVg_i \int \frac{d^3p}{(2\pi\hbar)^3} \left[ - f_i \ln f_i \mp (1 \mp f_i)\ln( 1 \mp f_i)  \right]
\end{eqnarray*}
Integrating through the angular parts, one obtains the entropy density ($\hbar=k=1$):
\begin{equation}
\label{Entropy}
s_i = \frac{g_i}{2\pi^2} \int_0^\infty p^2dp \left[ - f_i\ln f_i \mp (1 \mp f_i)\ln( 1 \mp f_i)  \right]
\end{equation}

\subsection{Change of variables}

The integrals for $n_i$ and $s_i$ are in $p$, but for $p \gg m$:
\begin{equation}
f_i(p) =
        \left[e^{\frac{\sqrt{p^2+m_i^2}-\mu_i}T} \pm 1 \right]^{-1}
\approx \left[e^{\frac{(p + m_i^2/2p_i) - \mu_i}T} \pm 1
\right]^{-1} \approx \left[e^{\frac{p - \mu_i}{T}} \right]^{-1} =
e^{-\frac{p - \mu_i}{T}}
\end{equation}
So the integrand is modulated by a decreasing exponential of form
$e^{-p/T}$.  This is not desirable, as one would like the
integrals to not be dominated by terms involving the temperature
(to be able to observe qualitative behaviour).  This is easily
fixed by introducing $x = p/T$, $\Rightarrow$ $p^2dp = T^3 x^2dx$,
and the limits of integration remain $0$ to $\infty$.  The two
densities are then:
\begin{eqnarray}
\label{IntX}
n_i &= &T^3 \frac {g_i} {2\pi^2} \int_0^\infty f_i(x) x^2 dx     \\
s_i &= &T^3 \frac {g_i} {2\pi^2} \int_0^\infty \left[ - f_i(x) \ln f_i(x) \mp (1 \mp f_i(x))\ln( 1 \mp f_i(x)) \right] x^2dx \notag
\end{eqnarray}
where we now have:
\begin{equation}
f_i(x) =  \frac 1 {e^{\sqrt{x^2+(m_i/T)^2}-\mu_i/T} \pm 1 }
\end{equation}
The quantities $m_i/T$ and $\mu_i/T$ are constants for each
particle and parameter set.


\section{Gauss-Laguerre Integration Technique}

The above integrals can be evaluated using a standard Simpson
method of dividing the domain into strips and approximating the
function on each strip by a parabola; the total integral being the
sum of the areas of all the strips.  This method is very accurate
(provided the strip widths are small), but also time-consuming.  I
did implement this method, integrating the domain from 0 to 20,
with $\Delta x = 0.002$.  However there exists a more efficient
and far more subtle integration technique.

\subsection{Gaussian quadratures}

The method of Gaussian quadratures is not a very well known method
of calculating integrals, as it is only readily applicable to a
limited set of integrands.  This topic can be found in the
literature, although the proofs of relevant theorems are not
presented, and I've been unable to reproduce them.  The following
arguments are taken from \emph{Numerical Recipes} \cite{Recipes},
combined with my own working.

Consider the class of functions that take the form of a polynomial
multiplied by some known function $W(x)$.  Then given the function
$W(x)$ and some integer $N$, one can create a set of $N$ weights
$w_i$ and abscissas $x_i$, to approximate the integral by a finite
summation:
\begin{equation}
\label{Approx}
\int_a^b W(x) f(x)dx \approx \sum_{i=1}^{N} w_i f(x_i)
\end{equation}
such that the approximation is exact if $f(x)$ is a polynomial.
Now for a given $N$, there are $2N$ parameters that can be chosen
at will (the weights and abscissas).  It then follows that
(\ref{Approx}) can only be exact for $2N$ linearly independent
polynomials, and it is usually chosen such that it holds for all
$f(x)$ with degree $D \leq 2N-1$.  For such a choice,
approximating a more general integral by a finite sum is only a
accurate if $f(x)$ can be ``well approximated by a polynomial''.
However almost all well-behaved (infinitely-differentiable)
functions can be well approximated by a polynomial of sufficiently
high degree. The value of $N$ ought to be chosen sufficiently
large to accommodate for slightly more complicated functions.

\subsubsection{Orthogonal polynomials}

For a specified function $W(x)$ and limits of integration, define
a ``scalar product'':
\begin{equation}
\left<f|g\right> \equiv \int_a^b W(x)f(x)g(x) dx
\end{equation}
Then one can find a set of polynomials that satisfy (i) there
exists exactly one polynomial of order $j$, called $(p_j(x))$, and
(ii) all the polynomials are mutually orthogonal over the
specified weight function $W(x)$.  As a further requirement one
can require that the polynomials $p_j(x)$ be normalized -- the
scalar product with themselves giving unity.  Then:
\begin{equation}
\left< p_i(x) | p_j(x) \right> = \delta_{ij}
\end{equation}
A set of such orthonormal polynomials can constructed through a
recurrence relation (Grand-Schmidt procedure), although it may not
be the most efficient way.

It can be shown that the polynomial $p_j(x)$ has exactly $j$
distinct roots in the interval $(a,b)$.  This becomes relevant
when one uses the fundamental theorem of Gaussian quadratures: The
abscissas $x_i$ of the $N$-point Gaussian quadrature formula
$(\ref{Approx})$ with weighting function $W(x)$ on $(a,b)$ are
precisely the roots of the orthogonal polynomial $p_N(x)$.

Once the abscissas $x_1...x_n$ are known, the weights $w_i$ can be
found by solving:
\begin{equation}
\label{Ortho}
\sum_{i=1}^N p_k(x_i) w_i = \int_a^b W(x)p_k(x)dx = \int_a^b
W(x)p_k(x) dx \cdot \delta_{0k}
\end{equation}
for each value of $k$ from $0$ to $N-1$ (all integrands except for
$k=0$ are zero, due to orthogonality).  The parameters $x_i$ and
$w_i$ will make (\ref{Approx}) exact for all any function that can
be written as a linear combination of $p_i(x)$.

\newpage

\subsection{Laguerre polynomials}

The form of integrals that are useful for this problem are
polynomials modulated by $e^{-x}$, integrated over the positive
real axis.  For this purpose take $W(x) = e^{-x}$, and the limits
of integration from $0$ to $\infty$.  The orthogonal polynomials
can be defined in terms of a recurrence relation, although there
is a simpler method:
\begin{itemize}
\item Define $L_0(x) \equiv 1$.  Then $\int_0^\infty L_0(x)W(x) = 1$ as required.
\item For all other $n$ define $L_n \equiv \frac {e^x}{n!} \frac{d}{dx^n}(x^ne^{-x})$, from which one gets:
\begin{equation*}
\int_0^\infty L_n(x) W(x) dx = \int_0^\infty \frac 1{n!}
\frac{d}{dx^n}(x^n e^{-x}) dx = \frac 1 {n!} \left[
\frac{d}{dx^{n-1}}(x^ne^{-x}) \right]_0^\infty = 0
\end{equation*}
\end{itemize}

Directly showing that (\ref{Ortho}) is satisfied, and also giving
the first orthonormality condition (as $L_0 = 1$), if one rewrites
the equation as $\int_0^\infty L_0(x)L_n(x)W(x) = \delta_{0n}$.

To see the form of the Laguerre polynomials, the first four have
been calculated:
\begin{eqnarray*}
L_0(x) &= &1 \\
L_1(x) &= &1-x \\
L_2(x) &= &\tfrac 12(2-4x+x^2) \\
L_3(x) &= &\tfrac 16(6-18x+9x^2-x^3)
\end{eqnarray*}
A general formula can be guessed, and shown to be true.  To find
the coefficient of $x^k$ from the expression $\frac
d{dx^n}(x^ne^{-x})$, the product rule is used repetitively, and
the derivative must be applied $k$ times to the exponential and
$n-k$ times to the monomial.  This produces a factor of $(-1)^k
\cdot\frac{n!}{k!}$.  As there are $n$ derivatives applied in
total, and $k$ are to the exponential, there is an additional
factor of ${ n \choose k}$.  Hence:
\begin{equation}
\label{Poly}
L_n(x) = \frac{1}{n!} \sum_{k=0}^n (-1)^k \frac{n!}{k!} {n\choose k} x^k = \sum_{k=0}^n (-1)^k \frac{n!}{(n-k)!(k!)^2} x^k
\end{equation}

\subsubsection{Orthonormality}

The orthonormality of Laguerre polynomials is mentioned in all the
literature, although I have been unable to find a proper proof of
this property.  The normalization $\int_0^\infty L_n^2(x)W(x)dx=1$
can be derived from (\ref{Poly}) by calculating the coefficient of
each $x^k$, and using the property $\int_0^\infty x^ke^{-x}dx=k!$.
This method is however very long and algebraically intensive, and
not shown in this project.  To show that distinct Laguerre
polynomials are orthogonal is a harder task, one that I've been
unable to complete myself, or indeed find a proof of this
relation.

If one takes it for granted that Laguerre polynomials are
orthonormal, then one can expand any polynomial $f(x)$ of degree
less that $2N$ as a linear combination of $L_i(x)$, and as
integrals preserve associativity, (\ref{Approx}) will be exact for
all such $f(x)$.

\subsection{An explicit construction}

For $N=2$ it is possible to construct the abscissas and weights
directly, by simply assuming (\ref{Approx}) and $\int_0^\infty
x^ke^{-x}dx=k!$.  This gives a system of four equations:
\begin{eqnarray*}
      w_1 +      w_2 &= &1 \\
x_1   w_1 + x_2  w_2 &= &1 \\
x_1^2 w_1 + x_2^2w_2 &= &2 \\
x_1^3 w_1 + x_2^3w_2 &= &6
\end{eqnarray*}
Multiplying the second equation by $(x_1+x_2)$ yields:
\begin{equation*}
x_1 + x_2 = (x_1+x_2)(x_1w_1+x_2w_2) = x_1^2w_1 + x_1x_2(w_1+w_2)
+ x_2^2 w_2 = 2 + x_1x_2
\end{equation*}
While multiplying the third equation by $(x_1+x_2)$ yields:
\begin{equation*}
2(x_1 + x_2) = (x_1+x_2)(x_1^2w_1+x_2^2w_2) = x_1^3w_1 +
x_1x_2(x_1w_1 + x_2w_2) + x_2^3w_2 = 6 + x_1x_2
\end{equation*}
giving two equations for $(x_1+x_2)$ and $x_1x_2$.  It follows
that $x_1x_2 = 2$ and $x_1+x_2=4$.  Solving for the individual
$x$'s (ordering is not defined -- take $x_1 < x_2$) gives
\begin{equation*}
x_1 = 2-\sqrt{2} \qquad \hbox{ and } \qquad x_2 = 2 + \sqrt{2}
\end{equation*}
Note that the $x_i$'s are roots of the quadratic $x^2-4x+2$, which
is (up to a constant factor) $L_2(x)$.  Hence the abscissae have
been shown to correspond to the zeroes of the relevant Laguerre
polynomial.

The problem is now essentially solved.  Putting the found values
for the $x_i$'s into the original system leaves a set of linear
equations for $w_i$'s, which can be trivially solved. The
solution:
\begin{equation*}
w_1 = \frac 14 (2 + \sqrt{2}) \qquad \hbox{ and }\qquad  w_2 =
\frac 14 (2-\sqrt{2})
\end{equation*}
A little algebra will show that the integral equations for
(\ref{Approx}) are satisfied for $x^k$ for $k=0,1,2,3$. Hence the
two Laguerre point problem is solved. This result, although simple
to derive algebraically, allows one to integrate (from zero to
infinity) any exponentially modulated polynomial up to cubic by
simply evaluating the function at two points.

A result not really evident for the above case is the following.
Suppose we use $N$ Laguerre points.  Then ANY polynomial $P(x)$
which has zeroes at the Laguerre points will have a zero integral.
This despite the extra $N-1$ supposed degrees of freedom.  An
extreme case is a polynomial of the form:
\begin{equation*}
P(x) = (x_N-x) \times \Pi_{i=1}^{N-1} (x-x_i)^2
\end{equation*}
which is positive for all $x \leq x_N$ (without loss of generality
$x_N > x_i$), and only negative once $x \geq x_N$, by which time
$P(x)e^{-x}$ is expected to be heavily suppressed. This is not
actually true: $P(x)$ grows as $x^{2N-1}$ at $x \sim x_N$, which
is much faster than the negative exponential at the same point
(for $N=15$, $x_N \approx 48 \gg e^{x_N/(2N)}$).

\newpage

\subsection{Applicability of Laguerre integration}

Unfortunately, this neat form of integration is only exactly
applicable to integrals of the form (2.10) - polynomials modulated
by a decreasing exponential.  However the integrals that are
needed for quantum statistics, are not like that (2.9).  For large
values of $x$, $f_i(x) \rightarrow e^{-x}$.  However, there is the
$\epsilon = \pm1$ factor, which creates slight problems for $x
\approx 0$, and also the $m_i/T$ factor which deforms the
exponential factor.

The method can however be salvaged by taking a "large" number of
points.  For $n=15$ (as used), the method works for any polynomial
up to degree $29$.  Although I have no analytic proof, it is
rather intuitive that the method should still give a good
approximate answer.  There are $15$ points where the function is
evaluated.  The integral is completely determined by the 15 data
points.  Now consider an exponentially modulated polynomial
function passing through the same 15 points.  There are still 14
degrees of polynomial freedom, by using which one can approximate
$f_i(x)$ very well over it's domain.

Numerical results for performing the integrals using the
Gauss-Laguerre and Simpson method (considered almost exact) show a
difference between the two methods of less than 5 ppm (parts per
million). For more complex functions (such as $s_i$), the Laguerre
method appears even more doomed, as $s_i$ involves terms
proportional to $(1\mp f_i)\ln(1\mp f_i)$, where the exponential
fall-off is not clear.

Consider the part of the expression for $x \gg \frac {m}T$.  Then
the function $f$ of form (2.9) can be reasonably well approximated
by: $f_i(x) \simeq  \left[ e^{x-\mu/T} \pm 1\right]^{-1} \ll 1$.
Then using the Taylor expansion for the natural logarithm
 $\ln(1+x) \approx x$, the entropy integrand becomes:
\begin{eqnarray}
- f_i\ln f_i \mp (1 \mp f_i)\ln( 1 \mp f_i)
&\approx &-f_i \ln f_i \mp (1 \mp f_i) (\mp f_i) \notag \\
&=       &-f_i \ln f_i + (1\mp f_i) f_i \notag \\
&=       & f_i ( 1 \mp f_i - \ln f_i) \notag \\
&\approx & f_i ( 1 - \ln f_i ) \notag \\
&\approx & (1 + (x-\mu/T))e^{\mu/T - x} \notag \\
&=       & e^{\mu/T} (x+1-\mu/T) e^{-x}
\end{eqnarray}
where I have approximated the distribution function by the
Maxwellian limit for large $x$.  So the entropy integrand,
although complicated for $x \ll m/T$, is  in the required form of
a polynomial modulated by an exponential for large $x$.  So
Laguerre integration can be applied here as well, although more
cautiously than before. The integrals remain accurate (for the
worst data set) to within a half of a percent.


\chapter{Entropy Calculations}

The derived equation (\ref{Entropy}) gives an exact value for the
entropy density of hadron~$i$.  To calculate the total entropy
density of the system, one is required to sum over the
contributions of all the hadrons.  To this end I was provided with
a list of all known hadrons composed of the $u$, $d$ and $s$
quarks that have a mass below $2.6 \hbox{GeV}$.  This contains all
the unflavoured and strange mesons, baryons and antibaryons.
Charmed particles were not included, but the program can be easily
extended to cater for these additional hadrons.  The particle
listing was presented in the following format (corresponding to
$\pi^+$, $\rho^0$, $\overline{K^0}$, $p$, $\Delta^{++})$
\begin{verbatim}
1. 0.140 -1. 0. 0. 1.
3. 0.770 -1. 0. 0. 0.
1. 0.498 -1. 1. 0. 0.
2. 0.938  1. 0. 1. 1.
4. 1.232  1. 0. 1. 2.
\end{verbatim}
The first column corresponds to the degeneracy factor $g_s =
2s+1$.  The second column gives the mass of the hadron in GeV. The
third column gives $\epsilon = \pm 1$, depending on whether the
particle obeys Fermi-Dirac or Bose-Einstein statistics.  The next
three columns give the strangeness $\texttt{S}$, the baryon number
$\texttt{B}$ and the charge $\texttt{Q}$ of the particle. These
three parameters will be used in the calculation of the particle's
chemical potential (\ref{Chemical}).  These are all the parameters
that are required to be able to work out the entropy. The input
file contained 358 hadrons.

The chemical potential depends on the strangeness, baryon number
and charge of each particle, as well as the values of  the
corresponding potentials.  Charge independence has been assumed
$(\mu_Q = 0)$, which essentially disregards the last column of
input.  The strangeness potential is fixed by (\ref{NonStrange}),
as the total strangeness of the system must be zero.

\newpage

This leaves two free parameters: temperature $T$ and baryonic
chemical potential $\mu_B$.  By analysis of past heavy ion
collisions, and applying the thermal model to the particle ratios
observed, one can fit the parameters $T$, $\mu_B$ and the other
potentials to get the best agreement with experiment.  These
values correspond to the system parameters at \emph{chemical}
freeze-out.
\begin{center}
\begin{figure}[h]
\label{TvUb}
\centering
\includegraphics[width=15.25cm, height = 10.8125cm]{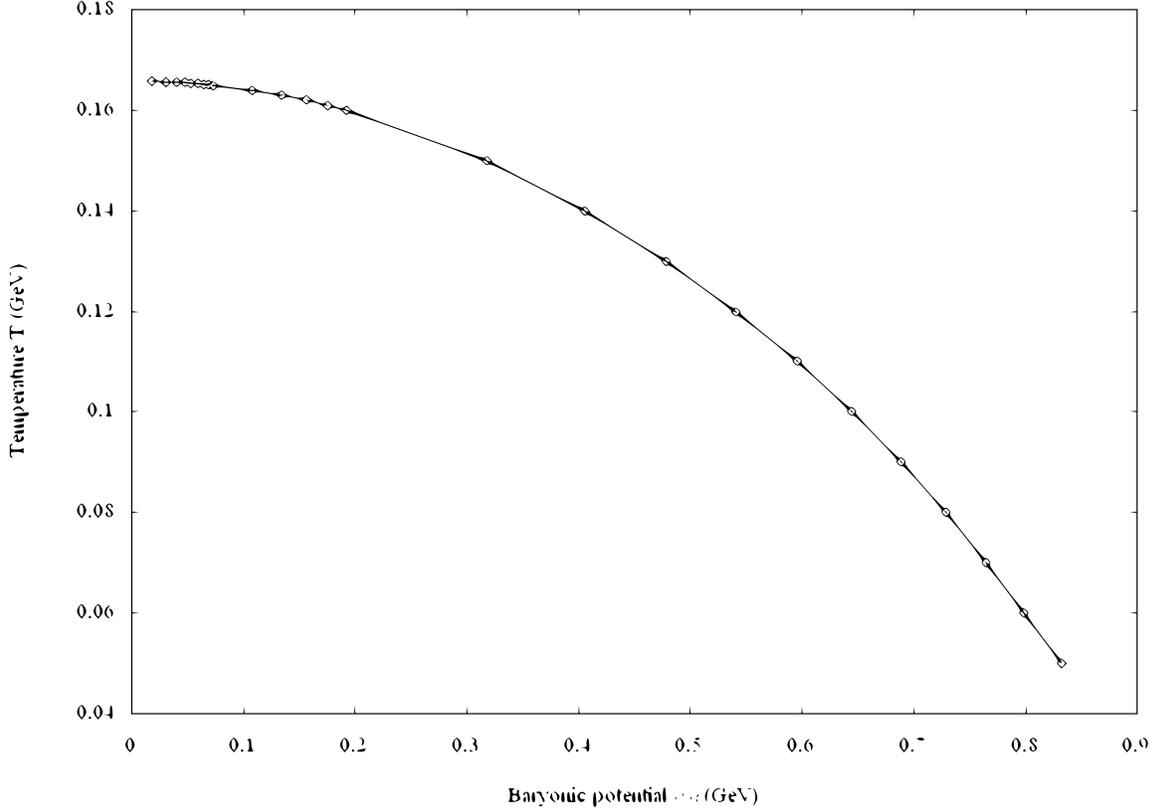}
\caption{Relationship between the temperature $T$ and chemical
potential $\mu_B$ at chemical freeze-out, as found by observing
heavy ion collisions}
\end{figure}
\end{center}
There are 25 data points shown in Fig 3.1, largely concentrated at
$T \approx 0.165\hbox{GeV}$.  The points furthest to the right
correspond to the lowest energy collisions, and as the beam
energies increase, the temperature $T$ at freeze-out increases,
while $\mu_B$ decreases. As the baryonic chemical potential can
never become negative (always excess baryons), and the temperature
at the highest energy collisions appears to tend to an almost
constant value, it is expected that when LHC becomes operational,
the parameters will have values $\mu_B \approx 0^+$, and $T
\approx 0.17\hbox{ GeV}$.

In all further calculations, the 25 data points describing the
freeze-out curve are used.  All plots shown are always against one
of $T$ or $\mu_B$, the other being implicit.

For all the points on the $T$-$\mu_B$ curve, it has been pointed
out by Cleymans and Redlich \cite{GeV} that the average energy per
hadron at the chemical freeze-out is almost constant throughout.
In particular $\frac \varepsilon n \approx 1$~GeV.

Further, it has been shown by Cleymans \cite{muB}, that the
baryonic potential can be fitted to the beam energy using a rather
simple formula:
\begin{equation}
\mu_B(s) \approx \frac{a}{1+\sqrt{s}/b}
\end{equation}
with $a\approx 1.27$ GeV, and $b\approx 4.3$ GeV.  The temperature
dependence on the beam energy can be implemented using $\frac
\varepsilon n \approx 1$~GeV.  Plots of the temperature and
baryonic potential as a function of $\sqrt{s}$ are provided in
\cite{muB} -- here I only reproduce the curve for $\mu_B$, due to
the unavailability of an analytic form of the $T$ dependence.
\begin{center}
\begin{figure}[h]
\centering
\includegraphics[width=15.25cm, height = 10.8125cm]{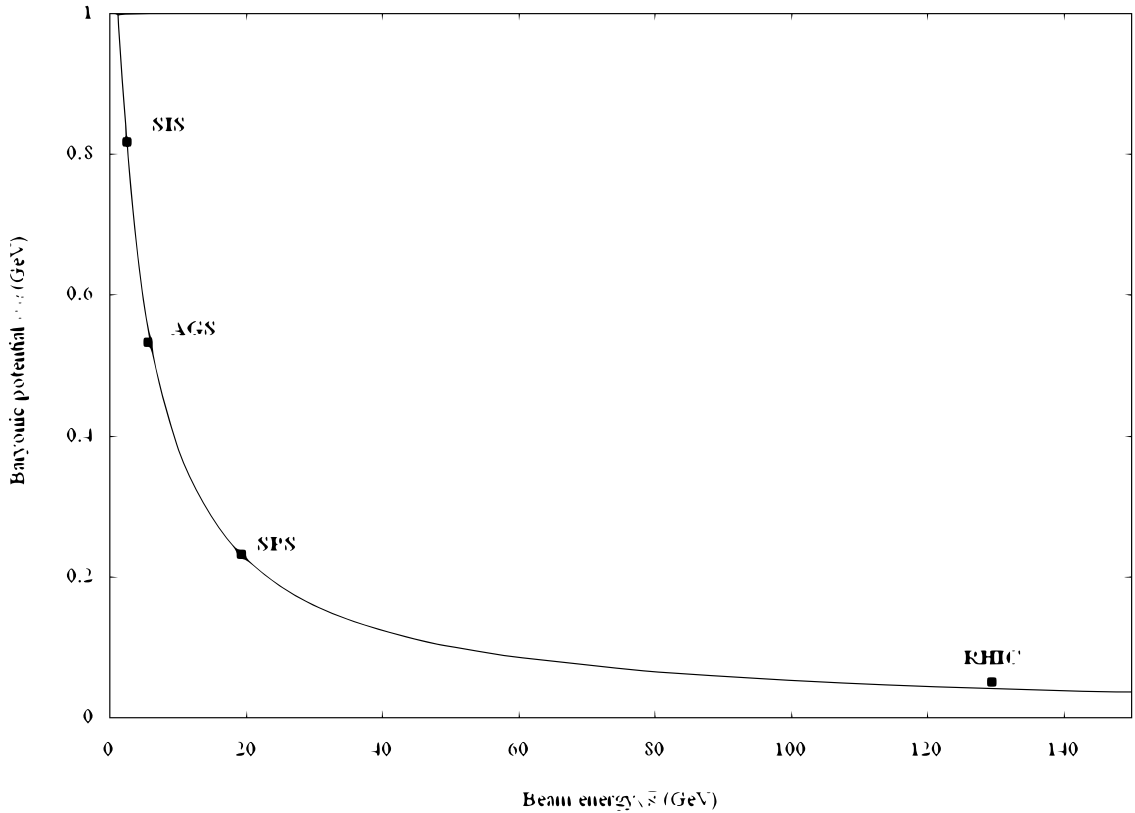}
\caption{Value of baryonic chemical potential $\mu_B$ as a
function of beam energy $\sqrt{s}$, with parameter results from
four experiments put it.}
\end{figure}
\end{center}
Various features of the thermal model are calculated (next
sections) in terms of $T$ and $\mu_B$, and to find the
corresponding beam energy $\sqrt{s}$ for a given $(T,\mu_B)$, the
above relationship will be used.

\newpage

\section{Determining $\mu_S$}

The strangeness chemical potential $\mu_S$ is uniquely defined by
$T$ and $\mu_B$, but there it however no analytic way of
determining it directly.  It needs to be solved numerically. To do
so, it is essential to first know approximately what value is
expected.

\subsection{First estimate}

Every strange particle has a corresponding anti-particle, which
has exactly opposite strangeness, while identical mass and
degeneracy factor.  They have opposite charges, but with $\mu_Q=0$
this is irrelevant. In the current discussion, a meson's chemical
potential depends only on the strangeness of the particle, so of
the two groups: strange mesons and anti-strange mesons, the one
with the greater chemical potential will exist in large
quantities.

For baryons however, there is $\mu_B$, which is always positive.
Hence there will always be more baryons than anti-baryons, and in
the baryon section of the fireball, there will be an excess of
strange quarks.

To keep total strangeness zero, there must be an excess of
anti-strange quarks in the mesons.  Hence $\mu_S \cdot
\texttt{S}(\overline{s}) > 0$.  As strangeness of $\overline{s}$
is $+1$, we deduce that $\mu_S > 0 $.

For an upper bound, I presumed that $\mu_S$ will scale not faster
than linearly with $\mu_B$, and have taken $\mu_S < \frac 13
\mu_B$. Although I have no analytical way of showing that this
will suffice, numerically it was shown to work.

Combining two bounds, the original interval was taken as: $0 <
\mu_S < \frac 13 \mu_B$.

\subsection{Numerics}

The momentum distribution function was originally mentioned in the
first chapter.  It is defined as:
\begin{equation}
\label{f_i}
f_i(p) = \left[e^{\frac{\sqrt{p^2+m_i^2}-\mu_i}{kT}} + \epsilon \right]^{-1}
\end{equation}
Then the number of particles of species $i$ that will be produced
is:
\begin{equation}
n_i = \frac {g_i} {2\pi^2} \int_0^\infty f_i(p) p^2 dp
= \frac {g_i} {2\pi^2} \int_0^\infty \frac {p^2 dp} {e^{\frac{E-\mu_i}{kT}}\pm 1}
\end{equation}

Combining this with (\ref{NonStrange}), for a given $T$ and
$\mu_B$:
\begin{equation}
\mathcal{F}(\mu_S) = \sum_i \texttt{S}_i n_i =  \sum_i \frac
{\texttt{S}_ig_i} {2\pi^2} \int_0^\infty \frac {p^2 dp}
{e^{\frac{E-\mu_i}{kT}}\pm 1} = 0
\end{equation}
The problem has been transformed into one of finding a zero of the
function $\mathcal{F}(\mu_S)$. Starting with the bounds $0 < \mu_S
< \frac 13\mu_B$, it turns out that the $\mathcal{F}$ is
monotonically growing over this interval $( \mathcal{F}(0) < 0$
and $\mathcal{F}(\frac 13\mu_B) > 0)$, and an iterative solution
(such as the bisection method) can be implemented to find a zero
of $\mathcal{F}$.

Importantly here, each evaluation of $\mathcal{F}$ requires the
evaluation of almost 200 integrals (one only needs to find $n_i$
for strange particles) of a non-standard form.  This can be done
using the Simpson method, but that is extremely time consuming,
and the program takes a very long time to find a value of $\mu_S$
to a sufficient accuracy.  I have calculated $n_i$ for a few
typical parameter sets, and have found that for the $\pi$'s the
Simpson and Laguerre methods differ by around 20 ppm, while for
all other particles the absolute difference was of order $10^{-8}$
or less. As pions are non-strange, all the necessary integrals can
be computed using the very fast Laguerre method (15 points), and
the bisection algorithm can be implemented.

\subsection{Results}

\begin{center}
\begin{figure}[h]
\centering
\includegraphics[width=15.25cm, height = 10.8125cm]{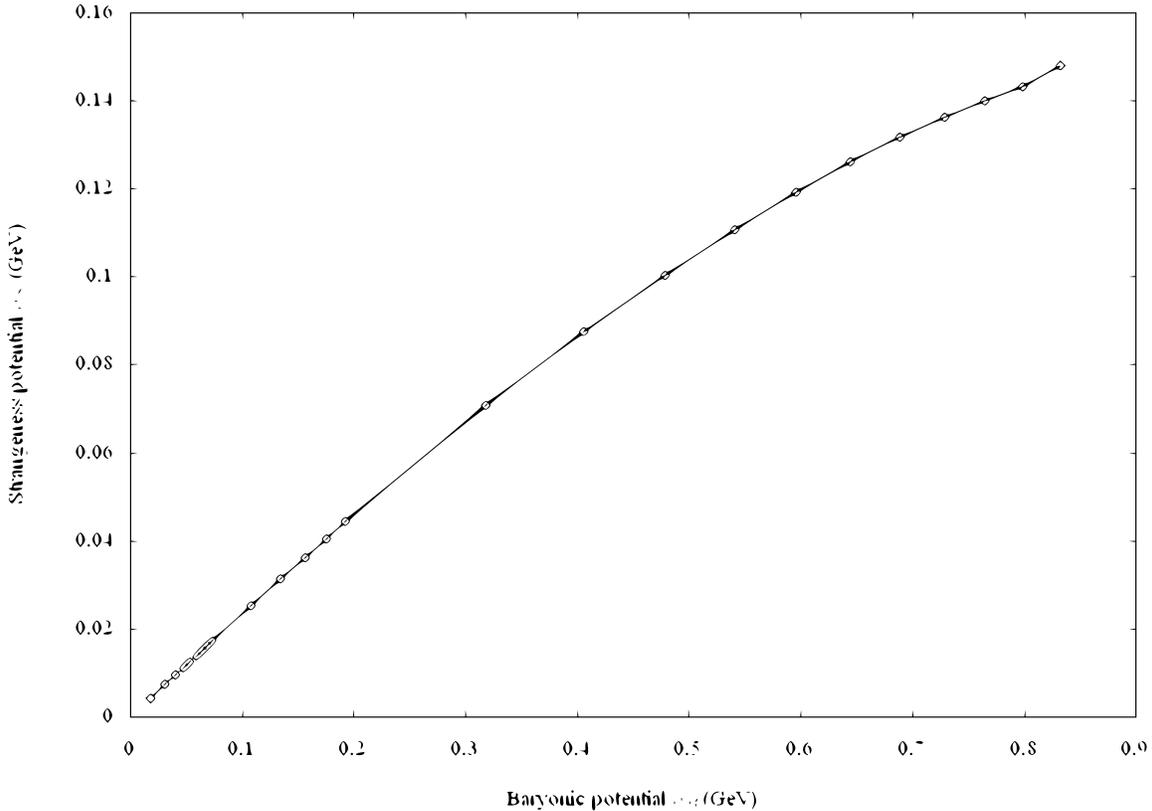}
\caption{Relationship between the strangeness $\mu_S$ and baryonic
potential $\mu_B$, calculated numerically along the freeze-out
curve}
\end{figure}
\end{center}

The result is rather neat, although not too surprising.  At very
low $\mu_B$, when the temperature is almost constant, the
strangeness potential goes linearly with $\mu_B$. As the baryonic
potential increases, so does $\mu_S$, but more slowly, and the
plot curves downwards.  The little twist at the last point is
rather unexpected, possibly due to uncertainty in $T$.
\begin{center}
\begin{figure}[h]
\centering
\includegraphics[width=15.25cm, height = 10.8125cm]{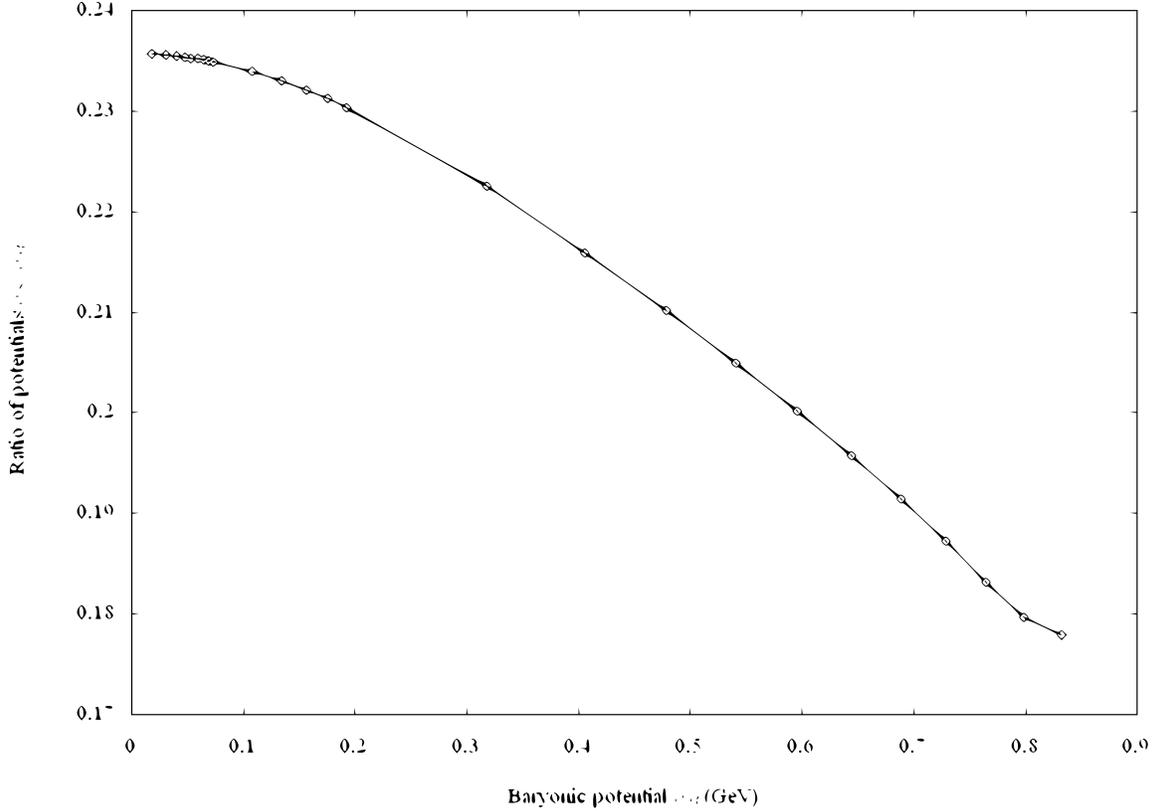}
\caption{Ratio of strangeness to baryonic potential
$(\mu_S/\mu_B)$ dependance on $\mu_B$}
\end{figure}
\end{center}
This plot is essentially the same as Fig 3.3, although it does
show more clearly the linear dependence of $\mu_S$ on $\mu_B$ as
the baryonic chemical potential is small, together with the
fall-off for larger values of $\mu_B$ (corresponding to lower
temperatures). The limiting value for the ratio of the chemical
potentials is:
\begin{equation}
\lim_{\mu_B \rightarrow 0} \frac {\mu_S}{\mu_B} \approx 0.236
\end{equation}
For a QGP, there are no hadrons, and for total strangeness to be
zero, $\mu(s) = \mu(\overline{s})$, and as the strange quark has
baryon number $\frac 13$ and strangeness $-1$ (opposites signs for
$\overline{s}$), it is required that $\mu_S = \frac 13 \mu_B$ in a
QGP.  The limiting value found does not correspond to that for a
QGP, as it is calculated for a hadron gas at freeze-out, which
differs from the hadron gas at formation (when QGP disassociates),
which in turn which differs from QGP through a phase transition,
breaking limit continuity.

\newpage

\section{Entropy in a Hadron Gas}

Once the strangeness potential $\mu_S$ has been determined, the
chemical potential for each hadron becomes known, and the
integrals can be computed.  Recalling that the entropy density for
one hadron (\ref{IntX}) can be written as:
\begin{eqnarray}
s_i &= &T^3 \frac {g_i} {2\pi^2} \int_0^\infty \left[ - f_i(x) \ln
f_i(x) \mp (1 \mp f_i(x))\ln( 1 \mp f_i(x)) \right] x^2dx \notag
\end{eqnarray}
it is expected that $s_i$ will eventually go as $T^3$ for
sufficiently large temperatures.  Also, as the $T^3$ term is
common for all hadrons at any given temperature, one can calculate
the integrals without it, and insert it later.  For this reason, I
define the \emph{normalized} entropy density for hadron $i$:
\begin{equation}
\mathcal{s}_i \equiv \frac {g_i}{2\pi^2} \int_0^\infty \left[ - f_i(x) \ln f_i(x) \mp (1 \mp f_i(x))\ln( 1 \mp f_i(x)) \right] x^2dx \notag
\end{equation}
Then the entropy density of each group of hadrons can be written
as:
\begin{eqnarray}
s_{\mathcal{M}} &= \hbox{ } {T^3} \mathcal{S_M} &= \hbox{ } {T^3} \sum_{Mesons} \mathcal{s}_i \notag \\
s_{\mathcal{B}} &= \hbox{ } {T^3} \mathcal{S_B} &= \hbox{ } {T^3} \sum_{Baryons} \mathcal{s}_i  \\
s_{\mathcal{T}} &= \hbox{ } {T^3} \mathcal{S_T} &= \hbox{ } {T^3} \sum_{Hadrons} \mathcal{s}_i \notag
\end{eqnarray}
which defines the group \emph{normalized} entropy densities
$\mathcal{S_M}$, $\mathcal{S_B}$ and $\mathcal{S_T}$.

\subsection{Meson-Baryon distribution}

The normalized entropy density $\mathcal{s}_i$ was evaluated for
each hadron.  This was done using both the Simpson and
Gauss-Laguerre integration techniques.  It was observed, that
although for several particles there was a ``large'' discrepancy
(up to 0.5\%)  between the two methods, these occurred only for
the heaviest particles, where $\sqrt{x^2 + \frac{m_i^2}{T^2}}
\approx x$ is a very bad approximation for all but very large $T$.
However, the hadron gas is always dominated by particles of lowest
mass.  In the case of mesons these are the $\pi$, $\eta$, $\rho$,
$K$'s. For baryons it is $p$, $n$ and light resonances, and for
all of the mentioned particles, the integration methods do not
differ by more that $10$ ppm.  If dealing with the (normalized)
entropy density for mesons, baryons and the total, the two methods
give very similar results (the relative differences never exceeded
20 ppm), which cannot be separated on a graph.

The $T$-$\mu_B$ relationship has been shown earlier $\hbox{(Fig
3.1}$), so all plots from here onwards will be against
temperature.

It is expected that for systems with low energy (and temperature),
the hadron gas will be dominated by the protons and neutrons
(baryons), and that the mesons will exist only in small
quantities.  Here, I put anti-baryons together with baryons.  For
very high temperatures (as observed at RHIC) the system is however
dominated by pions, and the entropy is expected to be mostly among
the mesons.  This is indeed observed:
\begin{center}
\begin{figure}[h]
\centering
\includegraphics[width=15.25cm, height = 10.8125cm]{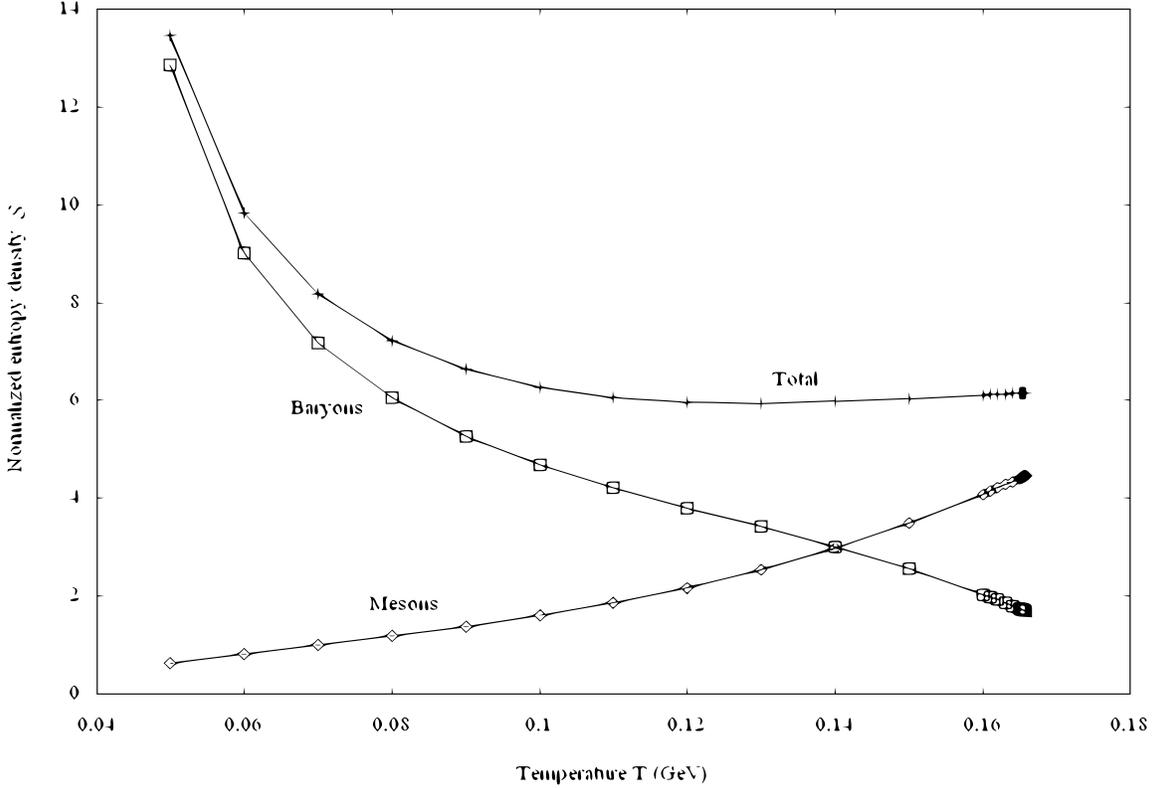}
\caption{Calculated normalized entropy density $\mathcal{S}$ for
mesons, baryons, and total.  The baryon and meson entropies are
seen to cross when $T \approx 140$MeV.}
\end{figure}
\end{center}
The plot contains many interesting features.  For low
temperatures, as was expected, the baryon entropy is far higher
than that for mesons.  Indeed, it appears to be far greater than
the entropy at higher temperatures - this is however not true. The
plot shows $s_i$ divided by the cube of the temperature.  The
baryon and meson curves intersect very near to $T=140$MeV.  This
corresponds to $\mu_B \approx 0.406$~GeV, and $\sqrt{s} \approx
9.3$ GeV.  This falls within the domain of the Alternating
Gradient System (AGS) at the Brookhaven National Lab.  For the
high temperatures (lots of points, as $T$ seems to approach a
constant value as $\mu_B \rightarrow 0$), the mesons dominate as
expected, while the baryons continue to contribute substantially
less for the increasing values of $T$.

The total normalized entropy density $\mathcal{S_T}$ exhibits a
very neat behaviour.  For $T < 100$MeV, it decreases almost
hyperbolically as the baryon contribution goes down.  It then
flattens out, has a minimum at $T = 130$MeV, and then increases
very slowly.  For the high $T$, the value of $\mathcal{S_T}$ is
almost constant, as if it was approaching a limit.  In that region
$\mathcal{S_T} \approx 6.15$

This limiting behaviour is not totally surprising.  It was
expected that as $T \gg m$ for the hadrons, $s_\mathcal{T} \sim T^3$, which
is the same as having $\mathcal{S_T}$ constant.  However this
limiting behaviour is observed already at $T \sim 130$MeV, which
corresponds to the pion mass, and is a lot less than the
masses of all the baryons.

For completeness, one may also look at a plot of the proper
entropy densities $s$ and their dependance on temperature:
\begin{center}
\begin{figure}[h]
\centering
\includegraphics[width=15.25cm, height = 10.8125cm]{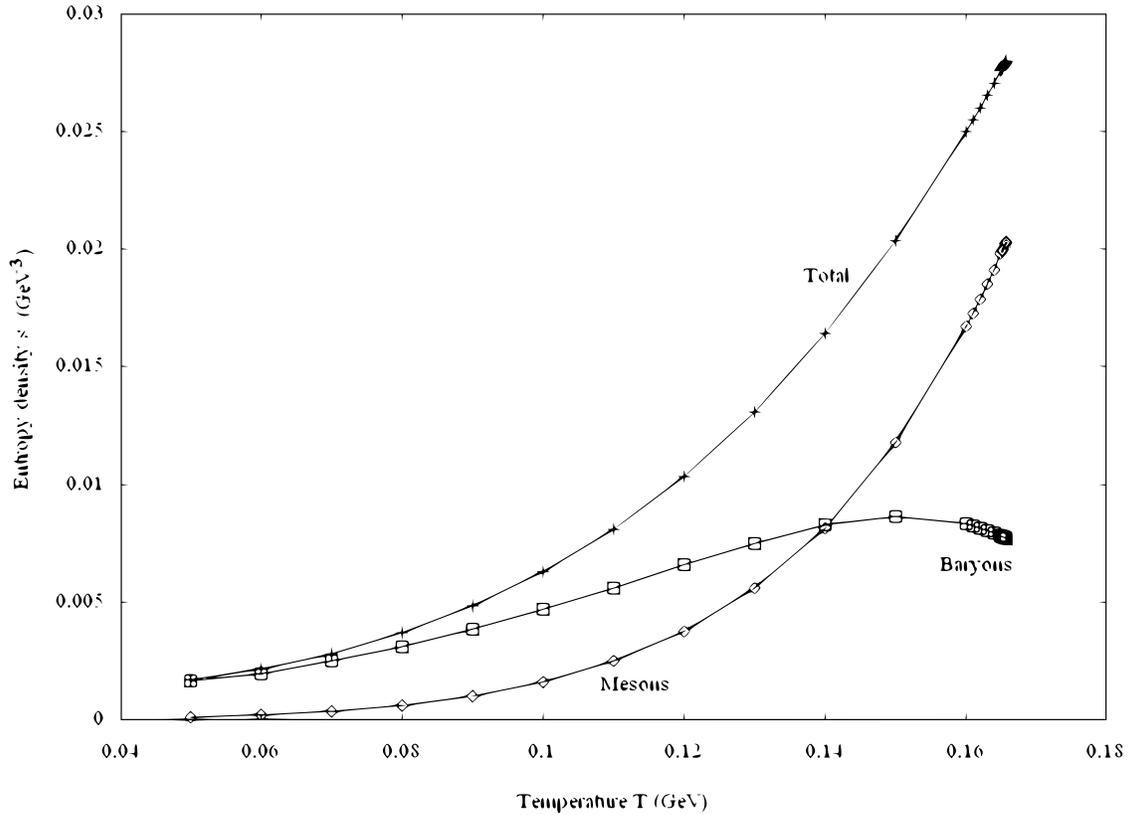}
\caption{Calculated entropy density $s$ for mesons, baryons, and
the total.}
\end{figure}
\end{center}
Here it can be clearly seen that the baryon entropy density, unlike $\mathcal{S_B}$ does
not decrease initially.  It increases (as one would expect),
although slower that $T^3$.  It does however reach a maximum at
$T=150$MeV, when $\mu_B$ starts decreasing fast relative to the
increase in $T$.  For mesons, the entropy density simply increases
very fast with $T$.  A power law fit gives $s_\mathcal{M} \sim T^{5.29 \pm
0.04}$.

\subsection{Baryon - Anti-baryon distribution}

Although this was not not originally planned, it was an automatic
consequence of doing the baryon-meson distributions, and though it
does not give any new results, it is a simple check on the
intuitive expectations for a hadronic gas.   At low temperatures,
baryons are favoured greatly over anti-baryons. At higher
temperatures (and beam energies), $\mu_B \rightarrow 0$, and
anti-particles are produced in sizeable quantities. Although there
is still a bias towards baryons, it is far smaller.

To do the plot, I originally planned to show $\mathcal{S_B}$ and
$\mathcal{S_{\overline{B}}}$, but that plot is not clear.  Recall
that $\mathcal{S}$ for baryons plus anti-baryons is maximal for
low $T$, and in that region $\mathcal{S_{\overline{B}}} \approx
0$.  Only at high $T$ will the anti-baryons contribute
significantly, and so I plotted the proper entropy densities $s_\mathcal{B}$
and $s_{\overline{\mathcal{B}}}$.
\begin{center}
\begin{figure}[h]
\centering
\includegraphics[width=15.25cm, height = 10.8125cm]{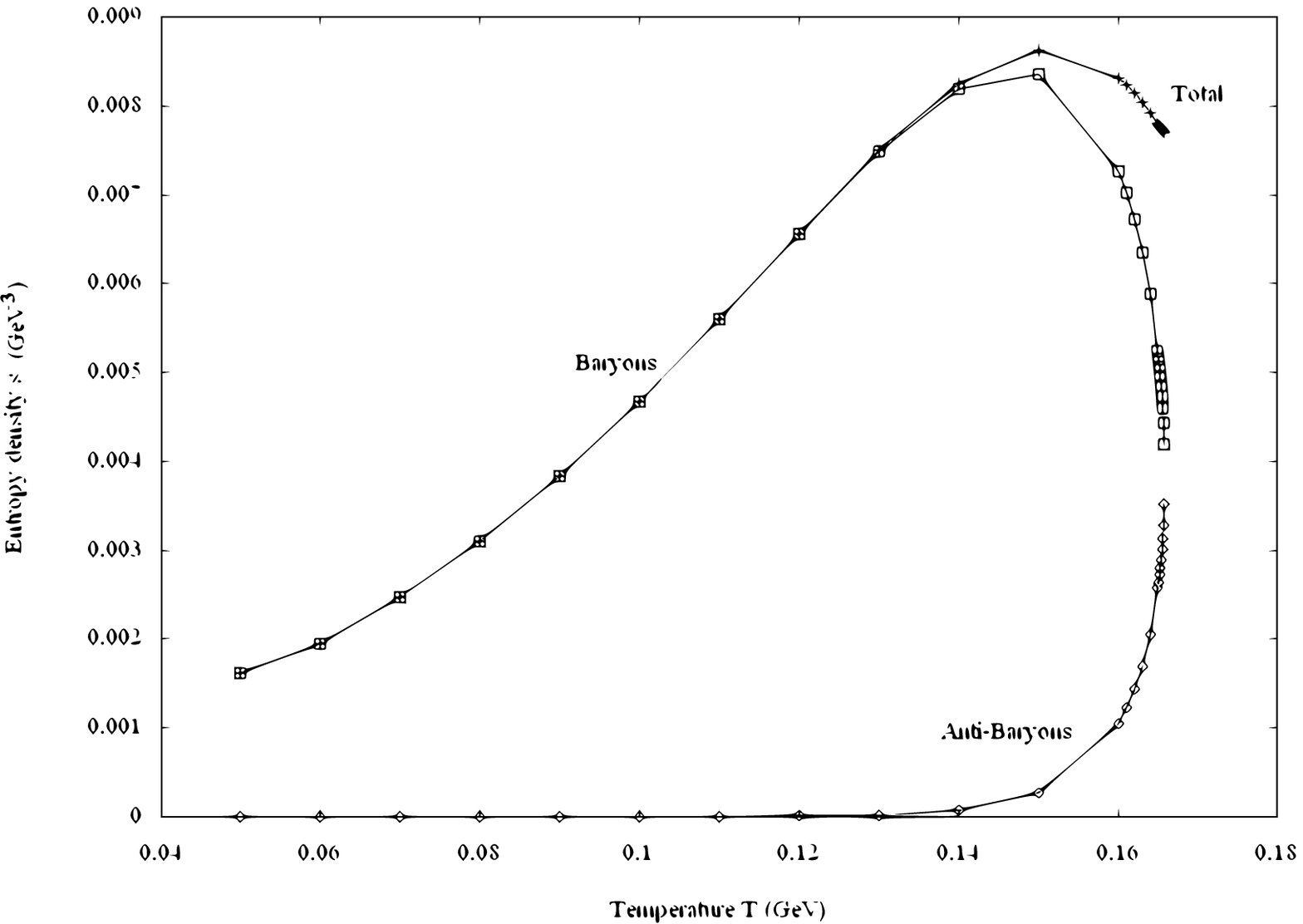}
\caption{Calculated entropy densities $s_\mathcal{B}$ and
$s_{\overline{\mathcal{B}}}$}
\end{figure}
\end{center}
Although maybe more pronounced at high $T$, the plot is as
expected.  The anti-baryons always contribute less than the
baryons, but as $\mu_B \rightarrow 0$, they do limit to the same
value as the baryons.  $s_\mathcal{B}$ drops sharply as $T \approx
165.5$ MeV remains almost constant while $\mu_B$ decays almost
linearly in the same region.

\subsection{Entropy of strange particles}

The distribution of the entropy among the mesons, baryons and
anti-baryons has been demonstrated.  I have mentioned that for
mesons, the entropy contributions come mainly from the $\pi$,
$\eta$ and $K$ particles, while baryon contributions come mainly
from nucleons and light resonances.  There are also lots of
strange particles with slightly higher masses, which for higher
$T$ are expected to contribute significantly.  This does indeed
happen:
\begin{center}
\begin{figure}[h]
\centering
\includegraphics[width=15.25cm, height = 10.8125cm]{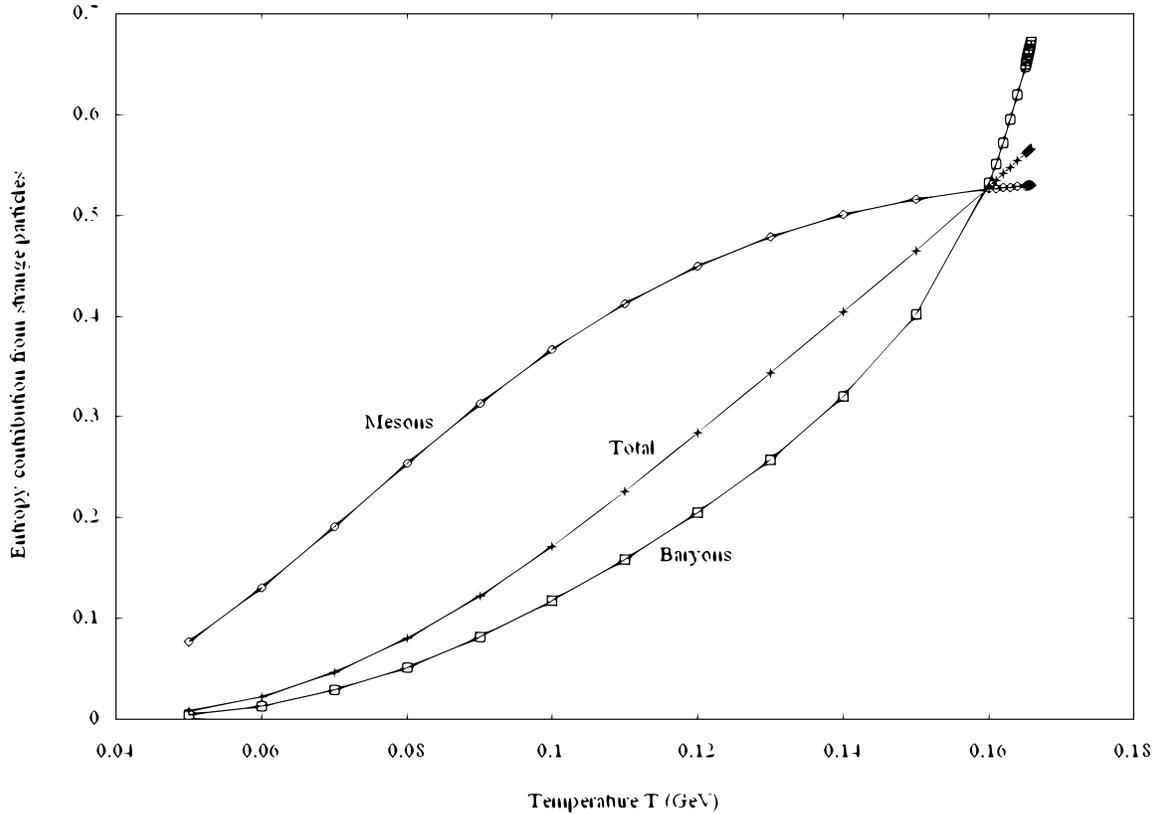}
\caption{Fractional contribution to the entropy of
mesons/baryons/total from strange particles (the $\phi =
s\overline{s}$ is considered unflavoured)}
\end{figure}
\end{center}
The result is rather interesting.  For the mesons, even at low
$T$, there is a little contribution from the kaons, and this
rises steadily with increasing $T$.  For the baryons, at low $T$
there are almost no strange particles (as expected), but for higher
$T$ this rises very sharply, and for $T>160$MeV, the strange
particles $(\Lambda, \Sigma \hbox{ etc.)}$ dominate.  This may be
accidental, but the meson and baryon contributions cross around
$0.5$, corresponding to equal strange and non-strange
contributions.


\section{Entropy in a QGP}

The calculations of entropy in a QGP are simpler than those in a
hadron gas, as there are far fewer ``particles''.  In this case, the
only contributors to the entropy are the quarks, anti-quarks and
gluons.  As for the hadrons, I only used unflavoured and strange
quarks ($u$, $d$, $s$ and the anti-quarks).  Gluons can be treated
in the same fashion.

For baryon numbers, one takes $\texttt{B}(q) = \frac 13$ and
$\texttt{B}(\overline{q}) = -\frac 13$.  Charge independence gives
$\mu_Q = 0$.  To ensure that strangeness is conserved, $\mu(s) =
\mu(\overline{s})$, and hence $\mu_S = \frac 13 \mu_B$.  This
deals with the three chemical potentials.

The $u$ and $d$ quarks are assumed massless, and the strange quark
was given a mass of $150$MeV.  Gluons are massless, and have no
charge, baryon number or strangeness, so $\mu(g) = 0$.  For spin
degeneracies $g_s$, I included colour, so $g_s(q)=6$ and
$g_s(g)=8\times 2 = 16$.  This is all the information required to
do the calculations.
\begin{center}
\begin{figure}[h]
\centering
\includegraphics[width=15.25cm, height = 10.8125cm]{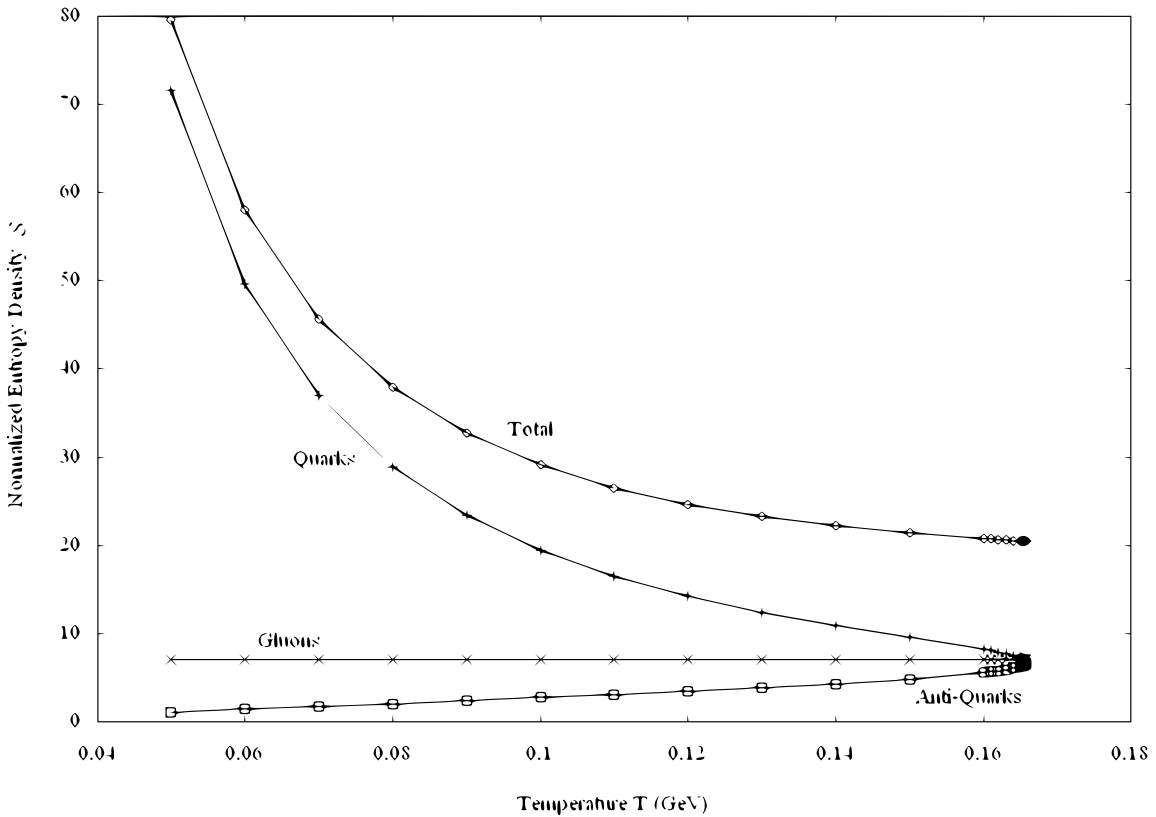}
\caption{Normalized entropy densities $\mathcal{S}$ for a QGP.  The
total is plotted, together with the individual contributions from
the quarks, anti-quarks and gluons}
\end{figure}
\end{center}
I had taken the $T$-$\mu_B$ set of parameters, and calculated the
entropy density that would result from a QGP at those values
(although $\mu_S$ is different).  The result:

As can be seen in Fig 3.9, the plot for the total entropy density
is very similar to that of the total entropy density of a hadronic
gas.  Starting very large, it decreases almost hyperbolically, and
then levels off at a value - presumably it will limit to a
constant.  There is however no little dip as in the HG case, and
for large $T$, while seemingly limiting to a contant, the
normalized density $\mathcal{S}$ is slowly decreasing for a QGP,
whereas it was slighly increasing in the case of a HG.  The
expected limiting value for the QGP being $\mathcal{S_{QGP}}
\approx 20.45$.

Next, it is observed that the gluon normalized entropy density
does not depend on temperature.  This is exactly what is expected,
as the gluons have no mass, and no chemical potential, so all the
$T$ factors drop out.  For $T \geq 165.5$MeV, I found that
$\mathcal{S}(g) > \mathcal{S}(q) > \mathcal{S}(\overline{q})$. It
is required that the quarks dominate the anti-quarks, but at high
temperatures the values become very close, and get exceeded by the
gluon entropy.  This was expected, although maybe only at higher
$T$.

The strangeness contribution to the total was also calculated:
\begin{center}
\begin{figure}[h]
\centering
\includegraphics[width=15.25cm, height = 10.8125cm]{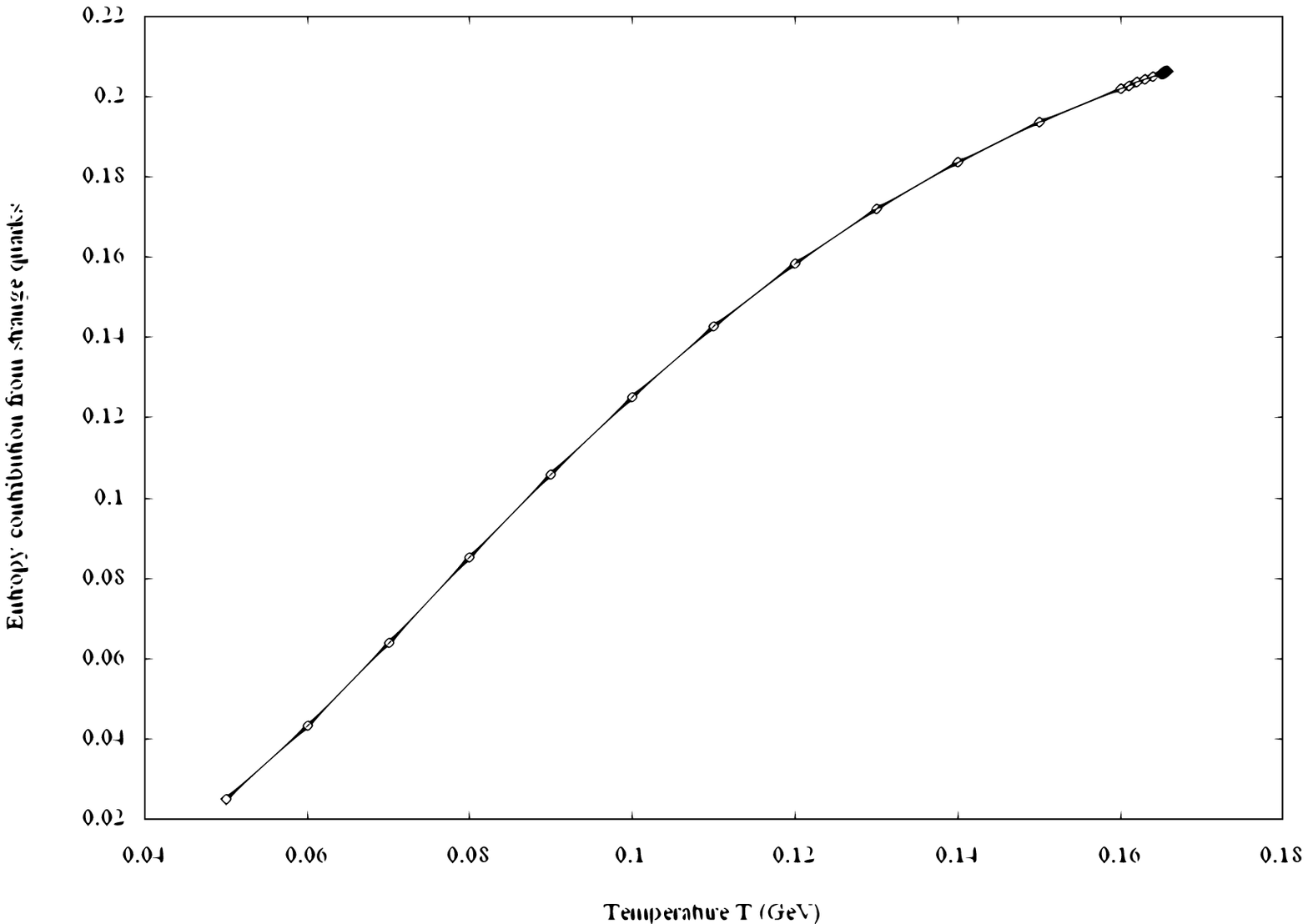}
\caption{Contribution of total entropy due to $s$ and
$\overline{s}$ quarks}
\end{figure}
\end{center}

The plot shown in Fig 3.10 does have a similar shape to the
strangeness contribution of mesons in a hadron gas.  There is
however a large difference in that for mesons, the strange
particle contribution went to a little more than a half, whereas
here the strange (anti-strange) quark contributions are never nore
than $0.21$ of the total. This can be easily explained -- for a
HG, the strange (anti-strange) quarks can bond with an $u$ or $d$
quark to form light kaons.  In QGP, strange quarks exist alone,
and as they are heavy ($m(s) \approx 150$MeV), their entropy will
be exceeded by the $u$ quarks, the $d$ quarks, AND the gluons at
high energies. So the upper bound for $s$, $\overline{s}$
contribution is $\frac 14 = 0.25$. Although this is not reached,
the contribution is in the expected range.

\section{Comparison of HG and QGP}

The limiting value for the normalized entropy density in a QGP has
been calculated to be a little more than three times greater than
the corresponding value in a hadron gas.  This is true
approximately true for all $T$.  A comparison plot for the total
entropy densities $s$ of the two phases is presented:
\begin{center}
\begin{figure}[h]
\centering
\includegraphics[width=15.25cm, height = 10.08125cm]{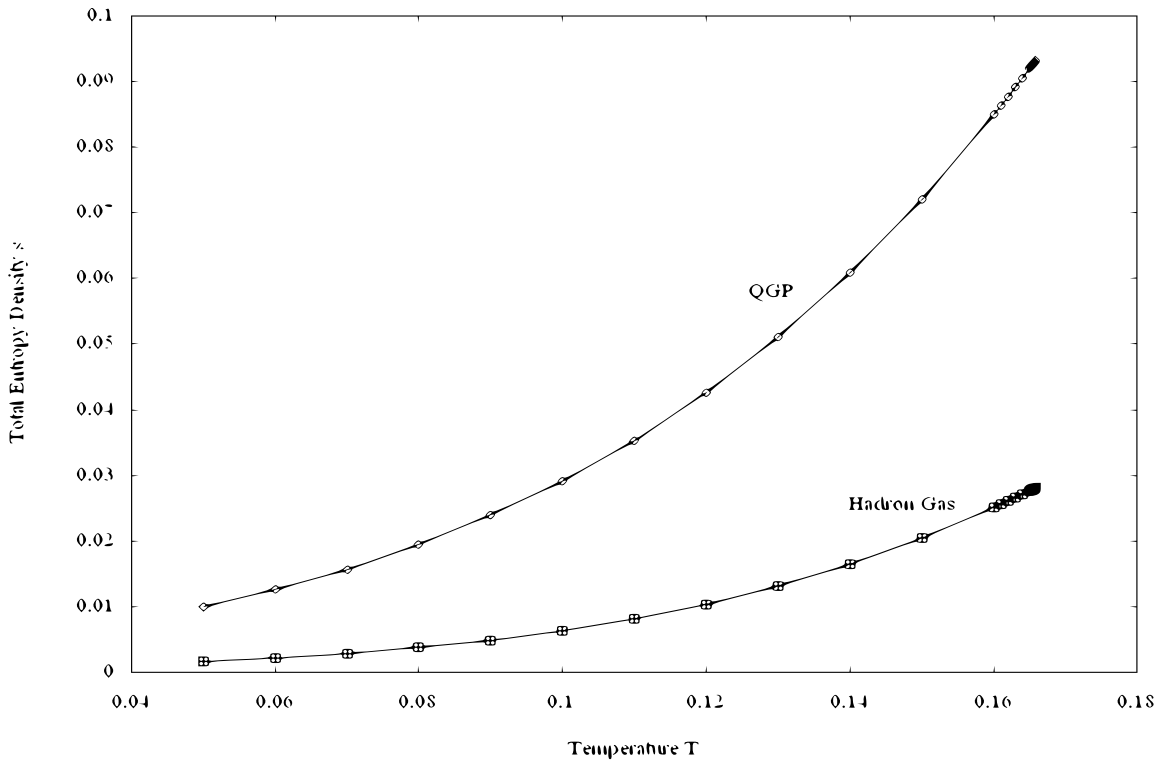}
\caption{Total entropy densities $s$ for a Hadron Gas and
 for a QGP}
\end{figure}
\end{center}
From Fig 3.11 one can clearly see that the entropy in a QGP is
higher by a factor of between 3 and 5 than the entropy in a hadron
gas with the same temperature and baryon chemical potential.  This
is due to the QGP consisting of essentially free quarks and
gluons, allowing for far greater degrees of freedom, and hence
directly increasing the entropy of the system.

It is expected that the transition between a hadron gas and a QGP
will be through a phase transition.  If one could take a hadron
gas and increase the temperature until a QGP formed, the entropy
would originally follow the HG curve, and at some critical
temperature ``jump'' to the QGP curve.  As the entropy curves
differ by a an almost constant factor $(\sim 3)$, a jump
discontinuity is expected, signifying a first-order phase
transition.

\section{Conclusion}

For an ideal hadron gas model of heavy ion collisions, the entropy
contributions have been calculated at freeze-out for a set of
parameter values satisfying the Cleymans-Redlich freeze-out
criterion. The entropy densities have been calculated for mesons
and baryons as well as the total (Fig 3.5 and Fig 3.6). It has
been demonstrated that at low temperatures (energies) the entropy
of the system is baryon dominated, while at high $T$ the mesons
dominate.  The baryon-meson cross-over occurs for $T\simeq 140$
MeV, which corresponds to $\sqrt{s} \approx 9.3$~GeV. This falls
within the domain of the AGS at Brookhaven.

For each of the considered parameter values $(T$ and $\mu_B)$ the
entropy density of a corresponding QGP has been calculated and
presented (Fig 3.11).  Due to ``free'' quarks in a QGP and extra
degrees of freedom, the entropy in each corresponding QGP is
significantly higher than that of the HG.  This shows that for a
transition from a hadron gas to a QGP, the entropy would have a
jump discontinuity, characteristic of a first order phase
transition.

At very low temperatures $T \approx 0.05$ GeV , the strange
particle contributions towards the entropy (in both HG and QGP)
are almost negligible.  As the energies increase, the strange
particle contributions increase monotonically with $T$, up to
$55\%$ in a hadron gas, and almost $25\%$ in a QGP.

Also, for a hadron gas the baryon -- anti-baryon dependence was
investigated.  At medium and low temperatures $(T \leq 140$ MeV),
the contributions from anti-baryons remain negligible.  Only at
very high energies (such as at RHIC), when $T \approx 165$ MeV,
the multiplicities of anti-baryons become significant, and the
ratio $\frac{\overline{p}}p \approx 0.8$ for the highest $T$ data
points.  For LHC it is expected that $\frac{\overline{p}}p
\rightarrow 1^-$.

\newpage
\LARGE
\begin{center}{\textbf{Acknowledgments\\}}\end{center}
\normalsize

I would like to thank my supervisor, Prof. Jean Cleymans, together
with the rest of the UCT-CERN group for their guidance,
encouragement and many helpful discussions throughout the duration
of this project. Extra thanks to Spencer Wheaton for proof-reading
this project so carefully.\\

Special thanks to Rudashan K. Chetty for his invaluable assistance
regarding my computer problems, and for ensuring I complete this
project. \\

To Tracy Timmins, thank you for all your support and encouragement
during the past months, which ultimately kept me sane.\\

I am grateful to the National Research Foundation (NRF) and the
UCT Honours Scholarship Committee for financial assistance.\\

Finally, to my parents, a special word of thanks for their
continual support and encouragement, without which I would never
have been able to do this project.

\end{document}